\documentclass[twocolumn,pre,showpacs]{revtex4}

\usepackage{graphicx}
\usepackage{epsfig}
\usepackage{dcolumn}
\usepackage{amsfonts}
\usepackage{amsmath}
\usepackage{amssymb}
\usepackage{bm}

\begin{document}


\newcommand{\brm}[1]{\bm{{\rm #1}}}

\title{On the relevance of percolation theory to the vulcanization transition}

\author{Hans-Karl Janssen}
\author{Olaf Stenull}
\affiliation{
Institut f\"{u}r Theoretische Physik 
III\\Heinrich-Heine-Universit\"{a}t\\Universit\"{a}tsstra{\ss}e 1\\
40225 D\"{u}sseldorf\\
Germany
}

\date{\today}

\begin{abstract}
The relationship between vulcanization and percolation is explored from the perspective of renormalized local field theory. We show rigorously that the vulcanization and percolation correlation functions are governed by the same Gell--Mann-Low renormalization group equation. Hence, all scaling aspects of the vulcanization transition are reigned by the critical exponents of the percolation universality class.
\end{abstract}

\maketitle

\noindent
\section{Introduction} 
Vulcanization~\cite{goldbart_castillo_zippelius_96&zippelius_goldbart_98} has a vast array of technological and commercial applications. The vulcanization process transforms a macromolecular liquid into an amorphous solid by randomly introducing permanent (chemical) crosslinks. In the liquid state all macromolecules are delocalized. In the amorphous solid state a non-zero fraction of macromolecules forms a macroscopic network. The constituents of this network are localized at random positions about which they execute thermal motion characterized by a distribution of finite localization lengths. The vulcanization transition (VT) between the two states (occurring at a critical density of crosslinks) represents a continuous phase transition. 

Due to the work of Goldbart, Zippelius, and coworkers (GZ)~\cite{goldbart&co_87to94} a rather comprehensive theoretical description of the VT exists to date on the level of a mean field approximation. The mean field theory gave a first glance on the relation between the VT and the percolation transition, and the critical exponents describing the VT were shown to be consistent with the mean field critical exponents for percolation.  Peng \textit{et al}.~\cite{peng&co_98} introduced a minimal model for the VT and discussed it under the aspects of universality as a common theoretical formulation of general amorphous solidification transitions (of which the VT is a prime example). Recently Peng and Goldbart (PG)~\cite{peng_goldbart_2000a} carried out a renormalization group improved one-loop calculation based on the minimal model~\cite{peng&co_98}. Their calculation showed that the critical exponents of vulcanization and percolation are in conformity to one-loop level.

In the past, the VT has often been addressed directly from the perspective of percolation theories~\cite{Perc}. In contrast to the work of GZ this perspective takes into account only a single ensemble of random connections, and therefore does not incorporate the effects of both quenched randomness and thermal fluctuations. Given that an essential aspect of the VT is the impact of the quenched random constraints on the thermal motion of the constituents, the a priori identification of the VT with percolation is a nontrivial matter and one may ask: What is the relevance of percolation theory to the VT? 

The purpose of this paper here is to explore the connection between vulcanization and percolation in depth. We compare the minimal model~\cite{peng&co_98} to a field theoretic model for the random resistor network (RRN) which we recently studied~\cite{stenull_janssen_oerding_99&stenull_2000}. Such a RRN is nothing more than a percolation model where randomly occupied bonds are assigned a finite nonzero conductance. In contrast to PG, which implement a ``momentum shell'' renormalization group, we use the powerful methods of renormalized local field theory. The momentum shell procedure, although physically intuitive, mixes scaling properties with features of other universal quantities like scaling functions. Therefore it is extremely difficult to extend the procedure beyond first order in perturbation theory. Renormalized field theory goes the other direction. Instead of the elimination of fluctuating degrees of freedom near the upper momentum cutoff and a rescaling of the momenta, the momentum cutoff is send to infinity at the beginning. The ultraviolet (UV) infinities resulting from this limiting procedure are eliminated by the so called renormalization factors. These renormalization factors then determine the Gell--Mann-Low renormalization group equation (RGE) which encodes all the scaling properties of the theory. This RGE represents the cleanest way to analyze the full scaling structure and its critical exponents. Physical properties beyond the scaling structure, however, cannot be inferred from the RGE. Indeed, these may be distinct for different theories like vulcanization and percolation. A further advantage of renormalized field theory is that it allows to find out general properties of the renormalization factors (and hence the RGE) to all orders of perturbation theory because they result solely from so called superficially divergent Feynman diagrams. Our analysis presented in this paper thrives on this advantage. We show rigorously that vulcanization and percolation involve the same primitive divergences and hence are governed by the same critical exponents. Furthermore, we compare the order parameters of the VT and the RRN.

\section{Modelling vulcanization and percolation}
The theory of GZ and PG is based on the order parameter field
\begin{equation}
\Omega (\hat{r})=\int_{\hat{k}}\widetilde{\Omega }(\hat{k}){\rm e}^{i\hat{k}
\cdot \hat{r}}
\end{equation}
defined on the replicated $d$-dimensional real space with coordinates $\hat{r}=({\bf r}_{1},\cdots ,{\bf r}_{n})$. The corresponding replicated wave vectors are $\hat{k}=({\bf k}_{1},\cdots ,{\bf k}_{n})$. Note that our $n$ corresponds to $(n+1)$ in the theory by GZ. The replica limit $n\to 1$ has to be taken before any other limit. The volume $V$ of the real space is considered as being finite. Thus, $\hat{k}$ is a discrete vector. $\int_{\hat{k}}\cdots$ is an abbreviation for $\sum_{\hat{k}}\cdots$ which is in the infinite volume limit equivalent to $(2\pi )^{-nd}\int d^{nd}\hat{k}\cdots$. Microscopically, the order parameter is a $n$-fold correlation function of density fluctuations and characterizes the amorphous state. The fluctuations about the average density them-self represent a non-critical stochastic variable which is excluded by the constraints
\begin{equation}
\Omega_{\alpha }\left({\bf r}^{(\alpha )}\right)=\int \prod_{\beta \neq \alpha
}d^{d}r^{(\beta )} \, \Omega (\hat{r})=0 \, ,\ \alpha =1,\cdots ,n\ .
\end{equation}
Hence, $\widetilde{\Omega }(\hat{k})$ is only non-zero if $\hat{k}$ belongs to the ``higher replica sector'' (HRS), that means if at least two distinct wave vectors ${\bf k}^{(\alpha )},{\bf k}^{(\beta )}$ of $\hat{k}$ are non-zero. The minimal model for the vulcanization transition~\cite{peng&co_98} is defined by
\begin{equation}
\mathcal{H}_{\text{VT}} = \int d^{nd}\hat{r}\Biggl\{\frac{\tau }{2}\Omega (\hat{r})^{2}+
\frac{1}{2}\left[ \hat{\nabla}\Omega (\hat{r}) \right]^{2}-\frac{g}{6}\Omega (\hat{r})^{3} \Biggr\}\   \label{VTHam}
\end{equation}
where $\tau - \tau_c$ measures the distance to the critical crosslink density at $\tau _{c}$. In a mean field approximation $\tau $ is positive in the liquid phase and the VT occurs at $\tau =\tau _{c}=0$. By virtue of the usual identification of $\Omega (\hat{r})$ with the microscopic density correlation, $g$ is a positive coupling constant. The Hamiltonian $\mathcal{H}_{\text{VT}}$ is complete in the renormalization group sense, i.e., it contains all relevant couplings and neglects all irrelevant ones.

We are going to compare the perturbation theory based on the minimal model $\mathcal{H}_{\text{VT}}$ to the perturbation theory of the field theoretic model for the RRN~\cite{stephen_78,harris_lubensky_87,stenull_janssen_oerding_99&stenull_2000}. This field theoretic model is based on an order parameter field $\phi ({\bf x},\vec{\theta})$ which lives on the $d$-dimensional real space denoted by the coordinates ${\bf x}$. The variable $\vec{\theta}$ denotes the $D$-fold replicated voltage at position ${\bf x}$. For regularization purposes, $\vec{\theta}=\vec{\nu}\Delta \theta $ takes discrete values on a $D$-dimensional torus, the replica space, i.e., $\vec{\nu}$ is chosen to be a $D$-dimensional vector with integer components $\nu^{(\alpha )}$ satisfying $-M<\nu _{\alpha }\leq M$ and $\nu^{(\alpha )} = \nu^{(\alpha )} \mod (2M)$. The order parameter field is restricted by the condition
\begin{equation}
\sum_{\vec{\theta}} \phi({\bf x},\vec{\theta})=0\ .  \label{ConstrRRN}
\end{equation}
It follows that the replica space Fourier transform $\tilde{\phi}({\bf x},\vec{\lambda})$ of the order parameter field, defined by $\phi ({\bf x},\vec{\theta})= (2M)^{-D}\sum_{\vec{\lambda}}\tilde{\phi}({\bf x},\vec{\lambda}) \exp ( i\vec{\lambda}\cdot \vec{\theta} )$, satisfies $\tilde{\phi}({\bf x},\vec{\lambda} = \vec{0}) =0$. Without exception we study the limit $D\rightarrow 0$, $M\rightarrow \infty $ of the replica space with $(2M)^{D}\rightarrow 1$ and $\Delta \theta =\theta_{0}/\sqrt{M} \to 0$. Here $\theta_{0}$ is a constant which sets the width of the voltage interval such that $[-\theta_{0} \sqrt{M} < \theta^{(\alpha )} \leq \theta_{0} \sqrt{M}]$. In the limit $D\rightarrow 0$, $M\rightarrow \infty$ the constant $\theta _{0}$ plays the role of a redundant scaling parameter, i.e., the theory is independent of its value. In the following we write $\sum_{\vec{\theta}}\cdots \approx (\Delta \theta )^{D}\sum_{\vec{\theta}}\cdots \approx \int d^{D}\theta \ldots =:\int_{\vec{\theta}}\ldots$, where the approximations become exact in the limit studied.

The field theoretic Hamiltonian for the RRN reads
\begin{eqnarray}
\label{hamiltonian}
{\mathcal{H}}_{\text{RRN}}
&=&\int d^{d}x\,\int_{\vec{\theta}}\bigg\{ \frac{\tau }{2}\phi({\bf x},\vec{\theta})^{2}+\frac{1}{2}\left[ \nabla \phi({\bf x},\vec{\theta}) \right]^{2}
\nonumber \\
&+&\frac{w}{2}\left[ \nabla_{\theta } \phi ({\bf x},\vec{\theta})\right]^{2}-\frac{g}{6} \phi({\bf x},\vec{\theta})^{3}\bigg\} \, .  
\end{eqnarray}
The parameter $\tau - \tau_c \sim (p_{c}-p)$ specifies the deviation of the occupation probability $p$ from the critical probability $p_{c}$. In mean field theory the percolation transition happens at $\tau = \tau_c =0$. $w$ is proportional to the resistance of the individual random bonds.

In the percolating phase, $\tau < \tau_c$, the mean order parameter is given by
\begin{eqnarray}
\left\langle \tilde{\phi} ({\bf x},\vec{\lambda}) \right\rangle_{{\mathcal{H}}_{\text{RRN}}} &=&\bigg\langle {\rm \exp }\bigg(-
\frac{\vec{\lambda}^{2}}{4}R_{\infty} ({\bf x}) \bigg)\bigg\rangle_C 
\nonumber \\
&=&\bigg\langle \chi _{\infty }({\bf x}) \, \exp \bigg(-\frac{\vec{\lambda}^{2}}{4}R_{\infty}({\bf x}) \bigg)\bigg\rangle_C
\nonumber \\
&=& P_{\infty}\bigg\langle {\rm \exp }\bigg( -\frac{\vec{\lambda}^{2}}{4}R_{\infty} ({\bf x})\bigg)\bigg\rangle_C^\prime \, .  \label{OrdPar}
\end{eqnarray}
Here $R_{\infty }({\bf x})$ is the (random) resistance between an arbitrary point ${\bf x}$ and infinity.  $\chi_{\infty }({\bf x})$ is an indicator function which is one if ${\bf x}$ is connected to infinity, i.e., if ${\bf x}$ belongs to a percolating infinite cluster, and zero otherwise. $\langle \cdots \rangle_C$ denotes the disorder average over all configurations of the diluted lattice. $\langle \cdots \rangle_C^\prime$ stands for disorder averaging conditional to the constraint that ${\bf x}$ belongs to an infinite cluster. Note that $P_{\infty } = \langle \chi_{\infty }({\bf x}) \rangle_C$ is the percolation probability, i.e., the order parameter for usual (purely geometric) percolation.

${\mathcal{H}}_{\text{RRN}}$ reduces for $w\rightarrow 0$ to the Hamiltonian for the $n=(2M)^{D}$-state Potts model with $n\rightarrow 1$ for $D\rightarrow 0$. This is important because the Potts model is known to describe percolation in this limit~\cite{kasteleyn_fortuin_69}. The connection between ${\mathcal{H}}_{\text{RRN}}$ and the Potts Hamiltonian becomes evident by relabeling the $n$ fields $\phi$ by an index $\alpha =1,\cdots ,n$: $\phi({\bf x},\vec{\theta})\rightarrow \phi_{\alpha }({\bf x})$. The constraint (\ref{ConstrRRN}) then reads $\sum_{\alpha }\phi_{\alpha }({\bf x})=0$. Taking the usual base $e_{i }^{(\alpha )}$, $i=1,\ldots ,n-1$, with $\sum_{\alpha }e_{i}^{(\alpha )}=0$, $\sum_{\alpha }e_{i}^{(\alpha )}e_{j}^{(\alpha )}=\delta _{i,j}$, $\sum_{i}e_{i}^{(\alpha )}e_{i}^{(\beta )}=\delta _{\alpha ,\beta }-1/n$, and upon defining $(n-1)$ independent fields $s$ by $\phi_{\alpha }({\bf x})=\sum_{i}e_{i}^{(\alpha )}s _{i}(
{\bf x})$, we get the Potts-Hamiltonian in the form
\begin{eqnarray}
\mathcal{H}_{\text{Potts}}&=&\int d^{d}x\,\bigg\{\sum_{i}\bigg[ \frac{\tau }{2} s_{i} ({\bf x})^{2}+\frac{1}{2}\left[ \nabla s_{i}({\bf x}) \right]^{2}\bigg]
\nonumber \\
&-&\frac{g}{6}\sum_{i,j,k}\lambda_{ijk} s_{i}({\bf x}) s_{j}({\bf x}) s_{k}({\bf x})\bigg\}\, ,
\end{eqnarray}
where $\lambda _{ijk}=\sum_{\alpha }e_{i}^{(\alpha )}e_{j}^{(\alpha
)}e_{k}^{(\alpha )}$ is the usual ``Potts tensor''. Note that the
introduction of the finite bond resistance $w\neq 0$ reduces the Potts symmetry, i.e., the full permutation symmetry of the $n$ fields $s$ belonging to the basic representation of the symmetry group $S_{n}$, to translation and rotation symmetry in the $D$-dimensional replica space.

\section{Renormalization and scaling}
\label{scalingProps}
We proceed with a renormalization group analysis of the VT by employing standard methods of renormalized field theory~\cite{amit_zinn-justin}. The principal diagrammatic elements are easily gathered from $\mathcal{H}_{\text{VT}}$. First, we have the three-leg vertex $g$. Because the corresponding interaction is cubic, $d_c =6$ is found to be the upper critical dimension of the VT. Second, it follows from the quadratic part of $\mathcal{H}_{\text{VT}}$ that the principal Gaussian propagator is given by\begin{eqnarray}
\label{propdecomp}
G^{\text{bold}} \big( \hat{k} \big) = G \big( \hat{k} \big) \Big\{ 1 - \sum_{\alpha =1}^n \delta_{\hat{k}, \brm{k}^{(\alpha)}\hat{e}^{(\alpha)}} + (n-1) \delta_{\hat{k},\hat{0}} \Big\} \, ,
\end{eqnarray}
where $G \big( \hat{k} \big) = (\tau + \hat{k}^2)^{-1}$. $\hat{e}^{(\alpha)}$ is a $n$-dimensional vector whose $\alpha$-th component is one and all other components are zero. Accordingly, $\brm{k}^{(\alpha)} = \hat{k}\cdot \hat{e}^{(\alpha)}$ is the component of $\hat{k}$ containing replica $\alpha$. The part of $G^{\text{bold}}\big( \hat{k} \big)$ embraced by the curly brackets ensures the constraint $\hat{k} \in \text{HRS}$. Equation~(\ref{propdecomp}) can be interpreted so that $G^{\text{bold}}\big( \hat{k} \big)$ decomposes into three generic parts one of which is proportional to $(n-1)$ and hence can be discarded from the onset. We call the remaining propagators $G^{\text{am}} ( \hat{k} ) = G ( \hat{k} )$ and $G^{\text{liq}}( \brm{k}^{(\alpha)} ) = G ( \hat{k})\delta_{\hat{k}, \brm{k}^{(\alpha)}\hat{e}^{(\alpha)}}$ amorphous and liquid propagators, respectively, because the HRS fields entering $G^{\text{am}} ( \hat{k} )$ are capable of diagnosing amorphous solidification in contrast to 1RS fields constituting the $G^{\text{liq}}( \brm{k}^{(\alpha)} )$. Moreover, we refer to $G^{\text{liq}}( \brm{k}^{(\alpha)} )$ as having color $\alpha$. Due to the propagator decomposition each principal diagram decomposes into a sum of amorphous diagrams consisting of amorphous and liquid propagators. Note that such an amorphous diagram may feature liquid propagators of just one or of several colors.

Now we are in the position to address the renormalization of the VT Hamiltonian. For this task it is sufficient to consider those Feynman diagrams contributing to the superficially divergent vertex functions for $n\to 1$. Thus, we restrict ourselves to one-particle irreducible diagrams with two or three amputated external legs. In the following we denote generic diagrams of this type by $\mathfrak{D}_2 (\hat{K})$ and $\mathfrak{D}_3 (\{\hat{K}\})$, respectively. $\hat{K}$ stands for an external wave vector and $\{\hat{K}\}$ is an abbreviation for $\{ \hat{K}_1 ,\hat{K}_2 ,\hat{K}_3 = - \hat{K}_1 - \hat{K}_2\}$. Once the decomposition of the bold diagrams into amorphous diagrams has been accomplished (i.e., the constraint $\hat{k} \in \text{HRS}$ is implemented) it is save to take the continuum limit in $\hat{k}$-space. Then we extract the divergences of these diagrams by expanding in a Taylor series in the external wave vectors, $\mathfrak{D}_2 (\hat{K}) = \mathfrak{D}_2 (\hat{0}) + \mathfrak{D}_2^\prime (\hat{0})\hat{K}^2$ and $\mathfrak{D}_3 (\{\hat{K}\}) = \mathfrak{D}_3 (\{\hat{0}\})$, where higher order terms are discarded since they are superficially convergent.

In the following we use without exception dimensional regularization. Thus, UV-divergencies appear as poles in the deviation $\varepsilon =6-d$ from $d_c$. These poles are eliminated from superficially divergent vertex functions by using the renormalization scheme
\begin{eqnarray}
\Omega &\rightarrow &\mathaccent"7017{\Omega }=Z^{1/2}\Omega \, ,\quad \tau\rightarrow \mathaccent"7017{\tau }=\mathaccent"7017{\tau }
_{c}+Z^{-1}Z_{\tau }\tau \ ,  \nonumber \\
g^{2} &\rightarrow &\mathaccent"7017{g}^{2}=A_{\varepsilon
}Z^{-3/2}Z_{u}u\mu ^{\varepsilon }\ ,  \label{RenVT}
\end{eqnarray}
where $A_{\varepsilon }$ is a suitably chosen amplitude, $\mu ^{-1}$ is a convenient length scale, and $u$ is a dimensionless version of the coupling constant. In the minimal renormalization procedure, i.e., dimensional regularization in conjunction with minimal subtraction, $\mathaccent"7017{\tau }_{c}$ is zero and the $\varepsilon $-poles are eliminated by $Z$-factors of the form
\begin{equation}
Z_{\ldots }=1+\sum_{m=1}^{\infty }\frac{Y_{\ldots }^{(m)}(u)}{\varepsilon
^{m}}\, .
\end{equation}
The $Y_{\ldots }^{(m)}(u)$ are expansions in the coupling constant $u$ beginning with the power $u^{m}$. A central theorem of renormalization theory, cf.~Ref.~\cite{amit_zinn-justin}, ensures that this procedure is suitable to eliminate the UV-divergencies from any vertex function order by order in perturbation theory. The $Y_{\ldots }^{(1)}(u)$ resulting from the primitive divergences determine the RGE entirely.

First we analyze $\mathfrak{D}_l (\{\hat{0}\})$, $l$ being any positive integer, under the assumption that it contains, apart from amorphous propagators, liquid propagators of just a single color, say $\alpha$. Obviously there are $n$ different realizations of the generic type represented by $\mathfrak{D}_l (\{\hat{0}\})$ corresponding to the $n$ different colors the liquid propagators can have. The sum over all these realizations appears in the decomposition of the pertaining bold diagram. However, all these realizations symbolize equivalent mathematical expressions since the internal wave vectors labeled by the colors are merely integration variables. Hence the overall contribution of this generic type to the primitive divergence of its bold diagram is  
\begin{eqnarray}
\sum_{\alpha =1}^n \mathfrak{D}_l (\{\hat{0}\}) = n \mathfrak{D}_l (\{\hat{0}\}) \stackrel{n\to 1}{\longrightarrow} \mathfrak{D}_l (\{\hat{0}\}) \, .
\end{eqnarray}
Next we assume that $\mathfrak{D}_l (\{\hat{0}\})$ contains liquid propagators of several colors $\alpha ,\beta , \cdots$. Then the sum over the realizations of this generic type is given by
\begin{eqnarray}
&&\sum_{\alpha =1}^n \sum_{\beta (\neq \alpha ) =1}^n \cdots \mathfrak{D}_l (\{\hat{0}\}) 
\nonumber \\
&& \, 
= n(n-1)\cdots \mathfrak{D}_l (\{\hat{0}\})  \, ,
\end{eqnarray}
i.e., it vanishes for $n\to 1$.

Now we turn to $\mathfrak{D}_2 (\hat{K})$ and assume that all featured liquid propagators are of the same color, $\alpha$ for sake. Since this color is distinguished, it is convenient to isolate it by setting $\hat{K} = \hat{K}_{\alpha} + \hat{e}^{(\alpha)} \brm{k}^{(\alpha)}$ with $\hat{K}_{\alpha} = \hat{K} - \hat{e}^{(\alpha)} \brm{k}^{(\alpha)}$. Then $\mathfrak{D}_2 (\hat{K})$ has the expansion $\mathfrak{D}_2 (\hat{K}) = \mathfrak{D}_2 (\hat{0}) + D_0 \hat{K}_{\alpha}^2 + D_1 \brm{k}^{(\alpha) 2} + \cdots$. It is important to realize that the expansion coefficients $D_{...}$ are independent of the choice of the color. Thus, the summation over all realizations of the generic type represented by $\mathfrak{D}_2 (\hat{K})$ gives
\begin{eqnarray}
\label{2Leg}
\sum_{\alpha =1}^n \mathfrak{D}_2 (\hat{K}) \stackrel{n\to 1}{\longrightarrow} \mathfrak{D}_2 (\hat{0}) + D_1 \hat{K}^2 + \cdots \, ,
\end{eqnarray}
where we have exploited that $\sum_\alpha \hat{K}_{\alpha}^2 = (n-1) \hat{K}^2$ vanishes for $n\to 1$.  The case of several colors $\alpha ,\beta , \cdots$ can be analyzed as above by isolating the distinguished colors. $\mathfrak{D}_2 (\hat{K})$ has the expansion $\mathfrak{D}_2 (\hat{K}) = \mathfrak{D}_2 (\hat{0}) + D_0 \hat{K}_{\alpha , \beta , ...}^2 + D_1 \brm{k}^{(\alpha) 2} + D_2 \brm{k}^{(\beta) 2} +\cdots$ with $\hat{K}_{\alpha , \beta , ...} = \hat{K} - \hat{e}^{(\alpha)} \brm{k}^{(\alpha)} - \hat{e}^{(\beta)} \brm{k}^{(\beta)} - \cdots$. Then it is straightforward to check that
\begin{eqnarray}
\sum_{\alpha =1}^n \sum_{\beta (\neq \alpha ) =1}^n \cdots \mathfrak{D}_2 (\hat{K}) \propto (n-1) \, ,
\end{eqnarray}
i.e., the contribution of the two-leg diagrams featuring liquid propagators of several colors vanishes for $n\to 1$.

To analyze the single color diagrams further, we write $G(\hat{k})$ in Schwinger parameterization, $G(\hat{k})=\int_{0}^{\infty }dt\,{\rm e}^{-(\tau +\hat{k}^{2})t}$. Suppose that $\mathfrak{D}_2 (\hat{K})$ comprises $P$ propagators, $P^{\text{am}}$ being amorphous and $P^{\text{liq}}$ being liquid with color $\alpha$. Assume that the diagram has $L$ loops, i.e., we have to integrate over $L$ independent combinations $\{\hat{q}_{i}\}^{\prime }$ of internal wave vectors $\{\hat{q}_{1,}\cdots ,\hat{q} _{P}\}$. This integration is of the form
\begin{eqnarray}
\label{IntegrVT}
&&\int_{\{\hat{q}_{i}\}^{\prime }}\exp \biggl\{-\sum_{i\in
P^{\text{am}}}t_{i}\hat{Q}_{\alpha ,i }^{2}-\sum_{i\in P}t_{i} \brm{q}_{i}^{(\alpha )2}\biggr\} 
\nonumber \\
&=&\big[
F(\{t\}_{P^{\text{am}}})\big]^{n-1}\exp
\bigl\{-C(\{t\}_{P^{\text{am}}})\hat{K}_{\alpha
}^{2}\bigr\}  
\nonumber \\
&\times& A(\{t\})\exp \bigl\{-B(\{t\})\brm{k}_{}^{(\alpha )2}\bigr\}\, ,
\end{eqnarray}
where $\hat{Q}_{\alpha ,i } = \hat{Q}_{i} - \hat{e}^{(\alpha)} \brm{q}^{(\alpha)}$. $\{t\}_{P^{am}}$ denotes the subset of the $\{t_{1,}\cdots ,t_{P}\}$ belonging to the amorphous propagators. By comparing Eqs.~(\ref{2Leg}) and (\ref{IntegrVT}) in the limit $n\rightarrow 1$ we see that $\mathfrak{D}_{2}(0)$ results from the integral $\int d\{t\}A(\{t\})\exp (-\tau \sum_{i}t_{i})$ and $D_{1}$ from $\int d\{t\}A(\{t\})B(\{t\})\exp (-\tau \sum_{i}t_{i})$. We conclude from Eq.~(\ref{IntegrVT}) that these parts can be extracted directly by replacing all amorphous and liquid propagators by elementary propagators of the type $(\tau + \brm{q}^{2})^{-1}$. Moreover, we can simplify $\hat{K}^{2}$ to $\brm{k}^{2}$ once the wave vector integration has been carried out. An analogous, yet simpler, reasoning applies to $\mathfrak{D}_{3}(\{0\})$ containing a single color $\alpha$. The case that $\mathfrak{D}_2 (\hat{K})$ and $\mathfrak{D}_{3}(\{0\})$ contain solely amorphous propagators, i.e., no colors at all, can be analyzed in a similar fashion.

The quint-essence of our considerations is that primitive divergences steaming from diagrams with multiple colors drop out in the limit $n\to 0$. Hence it is sufficient for calculating the critical exponents of the VT to work with an arbitrary single color. Further, after the decomposition into amorphous diagrams all the propagators can be identified with elementary propagators of the type $(\tau + \brm{q}^{2})^{-1}$. We will see in the following that the perturbation series resulting from this effective decomposition coincides with the diagrammatic expansion of the field theory of RRN based on the Hamiltonian $\mathcal{H}_{\text{RRN}}$ in the limits $D,w \rightarrow 0$.

The cubic interaction term in $\mathcal{H}_{\text{RRN}}$ leads to the three-leg vertex $g$. The principal propagator for the RRN is given by
\begin{eqnarray}
G^{\text{bold}} ( \brm{k}, \vec{\lambda} ) = G ( \brm{k}, \vec{\lambda} ) \{ 1 - \delta_{\vec{\lambda}, \vec{0}} \} \, ,
\end{eqnarray}
where $G (\brm{k}, \vec{\lambda}) = (\tau + \brm{k}^2 + w \vec{\lambda}^2)^{-1}$. Due to the factor $\{ 1 - \delta_{\vec{\lambda}, \vec{0}} \}$ which enforces the constraint $\vec{\lambda} \neq \vec{0}$ the principal propagator decomposes in a conducting part $G^{\text{cond}} ( \brm{k}, \vec{\lambda}) = G ( \brm{k}, \vec{\lambda})$ carrying replica currents and an insulating  part $G^{\text{ins}} ( \brm{k}) = G ( \brm{k}, \vec{\lambda})\delta_{\vec{\lambda}, \vec{0}}$ not carrying replica currents. Each principal diagram decomposes into a sum of conducting diagrams consisting of conducting and insulating propagators. As soon as this decomposition is accomplished it is save to switch to continuous replica currents.

Now it is important to realize the one-to-one correspondence between conducting and amorphous propagators as well as the one-to-one correspondence between insulating propagators and liquid propagators for $n\to 1$. As far as primitive divergencies are concerned, these one-to-one correspondences lead identical diagrammatic expansions (including the combinatorial fore-factors of the diagrams) for the VT and the RRN up to apparent distinctions in the propagators. Due to these distinctions a diagram $\mathfrak{D}_{2}( {\bf k},\vec{\lambda} )$ analogous to $\mathfrak{D}_{2}(\hat{K})$ (with $P^{\text{cond}}\cong P^{\text{am}})$ involves instead of Eq.~(\ref{IntegrVT}) the integrations
\begin{eqnarray}
\label{intRRN}
&& \int_{\{\brm{q}_{i},\vec{\kappa}_{i}\}^{\prime }}\exp
\biggl\{-w\sum_{i\in
P^{cond}}t_{i}\vec{\kappa}_{i}^{2}-\sum_{i\in P}t_{i}{\bf
q}_{i}^{2}\biggr\} 
\nonumber \\
&=&\big[ F(\{wt\}_{P^{\text{cond}}})\big]^{D}\exp
\bigl\{-C(\{t\}_{P^{\text{cond}}})w\vec{
\lambda}^{2}\bigr\}  
\nonumber \\
&\times& A(\{t\})\exp \bigl\{-B(\{t\})\brm{k}^{2}\bigr\} \, ,
\label{IntegrRRN}
\end{eqnarray}
with the same functions $A,B,C,F$ as in Eq.~(\ref{IntegrVT}). Equation~(\ref{intRRN}) leads for $D \to 0$ to
\begin{equation}
\mathfrak{D}_{2}(\brm{k},\vec{\lambda})\stackrel{D\rightarrow 0}{\longrightarrow
}\mathfrak{D}_{2}(0)+D_{0}w\vec{\lambda}^{2}+D_{1}\brm{k}^{2}+\cdots \, ,
\end{equation}
where the divergent coefficients $\mathfrak{D}_{2}(0)$ and $D_{1}$ are identical to those appearing in Eq.~(\ref{2Leg}). Here in the RRN, however, a further divergent coefficient $D_{0}$ arises. Thus, the renormalization scheme 
\begin{eqnarray}
s &\rightarrow &\mathaccent"7017{s}=Z^{1/2}s\, ,\quad g^{2}\rightarrow \mathaccent"7017{g}^{2}=A_{\varepsilon }Z^{-3}Z_{u}u\mu ^{\varepsilon }\, ,
\nonumber \\
w &\rightarrow &\mathaccent"7017{w}=Z^{-1}Z_{w}w\, ,\quad \tau \rightarrow \mathaccent"7017{\tau }=\mathaccent"7017{\tau }_{c}+Z^{-1}Z_{\tau }\tau
\label{RenRRN}
\end{eqnarray}
involves a further renormalization factor $Z_{w}$. The other factors, $Z,Z_{\tau },Z_{u}$, are identical to those in the renormalization scheme (\ref{RenVT}) because the superficially divergent parts of the two diagrammatic expansions coincide to any order for $w \to 0$. This leads to the conclusion that the renormalization factors appearing in the renormalization scheme for the VT are identical to those for percolation.

Due to the independence of the unrenormalized theory of the arbitrary length scale $\mu ^{-1}$ introduced by renormalization the correlation functions
\begin{eqnarray}
\mathaccent"7017G_{N}^{(\text{VT})}(\{\hat{r}\},\mathaccent"7017{\tau},\mathaccent"7017{g})=\left\langle \mathaccent"7017\Omega (\hat{r}_{1})\cdots \mathaccent"7017\Omega (\hat{r}_{N})\right\rangle_{\mathcal{H}_{\text{VT}}}
\end{eqnarray}
of the VT order parameter satisfy the identity
\begin{equation}
\mu \frac{\partial }{\partial \mu }\mathaccent"7017G_{N}^{(\text{VT})}=0\, .
\end{equation}
This identity translates with the help of the renormalization scheme (\ref{RenVT}) via the Gell--Mann-Low-Wilson functions
\begin{eqnarray}
\beta (u)&=&\left. \mu \frac{\partial u}{\partial \mu }\right| _{0}\, ,\nonumber \\
\kappa (u)&=&\left. \mu \frac{\partial \ln \tau }{\partial \mu }\right|
_{0}\, , \quad
\nonumber \\
\gamma (u)&=&\left. \mu \frac{\partial \ln Z}{\partial \mu } \right| _{0}\, ,
\end{eqnarray}
where the bare quantities are kept fixed while taking the derivatives, into the RGE
\begin{equation}
\biggl[\mu \frac{\partial }{\partial \mu }+\beta \frac{\partial }{\partial u}
+\kappa \tau \frac{\partial }{\partial \tau }+\frac{N}{2}\gamma \biggr] G_{N}^{(\text{VT})}(\{\hat{r}\},\tau ,u,\mu )=0\, .  \label{RGE}
\end{equation}
Since the functions $\beta ,\kappa ,\gamma $ are entirely determined by the renormalization factors, im particular by the $Y_{\ldots }^{(1)}(u)$, the RGE (\ref{RGE}) is exactly equal to the RGE for the correlation functions of percolation theory. The RGE determines the scaling structure of a field theory. Thus, the VT has the same critical exponents as percolation,
\begin{equation}
\nu =\big[ 2-\kappa (u_{\ast })\big]^{-1}\, ,\quad \eta =\gamma (u_{\ast })\, ,
\end{equation}
where $u_{\ast }$ is the infrared stable fixed point determined by $\beta (u_{\ast})=0$. The percolation exponents are known to third order $\varepsilon$~\cite{alcantara_80}. Introducing the correlation length $\xi \sim \left| \tau \right|^{-\nu }$ and the order parameter exponent $\beta =\nu (d-2+\eta )/2$, the asymptotic solution of the RGE (\ref{RGE}) can be written as
\begin{equation}
G_{N}^{(\text{VT})}(\{\hat{r}\},\tau )=\xi ^{-N\beta /\nu }F_{N}^{(\text{VT})}(\{\hat{r}
/\xi \})\, .
\end{equation}

The RRN differs from percolation as long as the bond-resistance $w$ is finite. The bond-resistance constitutes a further scaling variable and leads to an additional derivative $\zeta w\partial /\partial w$, where $\zeta (u) = \mu \frac{\partial \ln w }{\partial \mu } |_0$, in the corresponding RGE~\cite{stenull_janssen_oerding_99&stenull_2000}. Defining the resistance crossover exponent by $\phi =\nu [2-\zeta (u_{\ast })]$, the asymptotic solution for the correlation functions of RRN can be written as
\begin{eqnarray}
\label{asySol}
&& G_{N}^{(\text{RRN})}(\{{\bf x},\vec{\theta}\},\tau ,w) 
\nonumber \\
&&= \, \xi ^{-N\beta /\nu} F_{N}^{(\text{RRN})}(\{{\bf x}/\xi ,\vec{\theta}/\sqrt{w\xi ^{\phi /\nu }}\})\, .
\end{eqnarray} 

Although the VT and percolation are showing the same scaling behavior, the universal scaling functions $F_{N}^{(\text{VT})}$ and $\lim_{w\to 0}F_{N}^{(\text{RRN})}$ may be different for both theories. This is due to the fact that these functions are not entirely determined by the superficially divergent parts of the corresponding vertex functions. The universal scaling functions cannot be calculated from the RGE.

\section{Order parameters}
To explore the relation between VT and percolation further we now revisit the order parameters. We start with the order parameter for the RRN, Eq.~(\ref{OrdPar}). Above the percolation point we deduce from Eq.~(\ref{asySol}) and the constraint (\ref{ConstrRRN}) that
\begin{eqnarray}
&&M^{(\text{RRN})} (\vec{\theta},\tau ,w)= \left\langle \phi ({\bf x},\vec{\theta}) \right\rangle_{{\mathcal{H}}_{\text{RRN}}} =
P_{\infty }(\tau ) \bigg\{ \int_{0}^{\infty }dx
\nonumber \\
&&\times \,  p^{(\text{RRN})}(x){\rm \exp }\Big( -x\frac{\vec{\theta}^{2}}{w\xi ^{\phi /\nu }}\Big)-1\bigg\} \, .
\label{RRN-OP}
\end{eqnarray}
Here $P_{\infty }(\tau )\sim \left| \tau \right| ^{\beta }$ is the
percolation probability and $w\xi ^{\phi /\nu }p^{(\text{RRN})}(w\xi ^{\phi /\nu }\Sigma_{\infty })$ is the probability distribution of the conductance $\Sigma _{\infty }=1/R_{\infty }$ from an arbitrary point on the percolating cluster to infinity. The digit $1$ in the bracket denotes the limit of $1/n=(2M)^{-D}$ for $D\rightarrow 0$. In the $n$-state Potts model limit $w\rightarrow 0$, we retrieve from Eq.~(\ref{RRN-OP}) the Potts order parameter $M^{(\text{Potts})}(\vec{\theta},\tau )=P_{\infty }(\tau )(\delta _{\vec{\theta},0}-1)$.

The order parameter for the VT has an analogous form~\cite{goldbart_castillo_zippelius_96&zippelius_goldbart_98,goldbart&co_87to94,peng&co_98}
\begin{eqnarray}
&&M^{(\text{VT})}(\hat{r},\tau )=\left\langle \Omega (\hat{r}) \right\rangle_{{\mathcal{H}}_{\text{VT}}} =P_{\infty }(\tau ) \bigg\{ \int_{0}^{\infty }dx
\nonumber \\
&&\times \, 
p^{(\text{VT})}(x){\rm \exp }\Big(-x\frac{\brm{R}^2}{\xi ^{2}}\Big)-1\bigg\} \, ,
\label{VT-OP}
\end{eqnarray}
where $\brm{R}^2 = \sum_{\alpha =1}^{n}({\bf r}^{(\alpha )}-{\bf \tilde{r}})^{2}$ corresponds to the radius of gyration of the replicas about a center of mass at ${\bf \tilde{r}} = \frac{1}{n} \sum_{\alpha =1}^{n}{\bf r}^{(\alpha )}$. $\xi ^{2}p^{(\text{VT})}(\xi ^{2}\sigma )$ is the probability distribution of the inverse squares $\sigma$ of the localization lengths. Here, the digit $1$ in the bracket denotes the limit of $V^{-(n-1)}$ for $n\rightarrow 1$. In general the two distributions $p^{(\text{VT})}(x)$ and $p^{(\text{RRN})}(x)$ are different but universal. A mean field calculation leads in both cases (for the RRN see Appendix~\ref{reduct}) to $P_{\infty }(\tau )=2\left| \tau \right| /g$ and to the same integro-differential equation~\cite{goldbart_castillo_zippelius_96&zippelius_goldbart_98,goldbart&co_87to94,peng&co_98}
\begin{equation}
4\frac{d}{dx}\biggl(x^{2}p(x)\biggr)=p(x)-\int_{0}^{x}dx^{\prime
}\,p(x-x^{\prime })p(x^{\prime })\, ,
\label{DiffGl}
\end{equation}
with $p(0)=p(\infty )=0$.
It is interesting that the Laplace transform $\tilde{p}(z)=\int_{0}^{\infty}dx\,p(x)\exp (-zx)$ and the differential equation that follows from Eq.~(\ref{DiffGl}),
\begin{equation}
4z\tilde{p}^{\prime \prime }(z)=\Big(1-\tilde{p}(z)\Big)\tilde{p}(z)\, ,
\end{equation}
with $\tilde{p}(0)=1$ and $\tilde{p}(\infty )=0$, was already introduced many years ago by Stephen~\cite{stephen_78} and Stinchcombe~\cite{Sti74} in their mean field theory of RRN and determination of the conductivity of a Bethe lattice, respectively.

\section{Concluding remarks}
We showed that the primitive divergences occurring in percolation and vulcanization are identical to arbitrary order in perturbation theory. Consequently both transitions are governed by the same Gell--Mann-Low RGE. This RGE determines entirely the scaling structure of both transitions. Hence, the scaling behavior of physically analogous quantities in both transitions is identical. In particular, the VT is governed by the critical exponents of the percolation universality class. Quantities which are not completely determined by superficial divergences cannot be calculated from the Gell--Mann-Low RGE. These quantities, e.g. scaling functions, may be different in both theories.

Moreover, we compared the order parameter for the RRN and the VT. These two have an analogous form. The order parameter for the RRN involves a scaling function which incorporates the distribution of the conductance to infinity whereas the order parameter for the VT features the distribution of the inverse squares of localization length. In mean field approximation, the corresponding distribution functions turn out to be identical. To determine these distribution functions beyond mean field level is an interesting issue for future work.

We would like to emphasize that the analysis of scaling properties presented in Sec.~\ref{scalingProps} is hardly feasible without employing the methods of renormalized local field theory. These methods provide a clear cut discrimination between scaling properties and other universal quantities like scaling functions. Furthermore, these methods allow to restrict attention to superficially divergent diagrams which simplifies the analysis tremendously.

The intimate relationship between vulcanization and percolation seems plausible because macroscopic connectivity is the central issue in both systems. There are, on the other hand, striking distinctions between vulcanization and percolation. For example, common percolation models like the RRN live on some underlying lattice whereas vulcanization involves no lattice. Vulcanization as considered in this paper features an excluded volume interaction which is extraneous to percolation. A similarity between the excluded volume interaction and the lattice is, however, that both prevent overlap of the network constituents, monomers and bonds, respectively. Another obvious distinction is that the usual percolation involves a single ensemble, viz.\ the ensemble of the diluted lattice configurations, whereas vulcanization features two ensembles: one pertaining to the thermal degrees of freedom and one to the crosslink distribution. It turns out, though, that fluctuations of the crosslink distribution play a more important role than thermal fluctuations do, at least as far as the connectivity aspects of the VT are concerned.

\begin{acknowledgments}
We acknowledge support by the Sonderforschungsbereich 237 ``Unordnung und gro{\ss}e Fluktuationen'' of the Deutsche Forschungsgemeinschaft.
\end{acknowledgments}

\appendix
\section{Reduction of superficially divergent diagrams}
\label{reduct}
In this appendix we demonstrate the simplicity of determining the one-loop contributions to the renormalization of the VT model. Consider the
self-energy diagram constructed of the HRS (bold) propagators, Fig.~1.
\begin{figure}
\epsfig{file=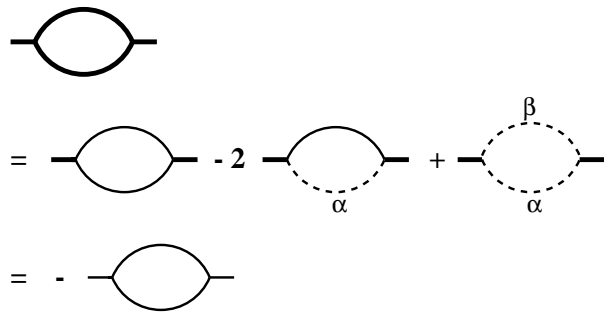,width=8.0cm}
\caption{Reduction of the one-loop self-energy diagram.}
\end{figure}
After decomposition in amorphous (thin) and liquid (dashed) propagators, we get the second line in Fig.~1, where summation over the colors $\alpha ,\beta $ is implied. We have shown that diagrams with different colors do not contribute to the primitive divergencies. Hence, we can set $\alpha =\beta $ in the third diagram. However, then the third diagram can be discarded because the wave vectors of the two propagators of the same color can not add to a HRS external wave vector at the vertices. Now we can replace all the propagators by the simple one $(\tau +{\bf q}_{i}^{2})^{-1}$ which results in the simple diagram appearing in the third line of Fig.~1. Therefore we get in dimensional regularization, after including the combinatorial factor $1/2$, the following one-loop contribution $\Gamma_{2}^{1-\text{loop}}$ to the two-leg vertex function $\Gamma_{2}$ (note that the vertex functions $\Gamma_{N}$ are defined as the negative of the corresponding diagrams):
\begin{eqnarray}
\Gamma_{2}^{1-\text{loop}}({\bf k}) &=& \frac{g^{2}}{2}\int_{{\bf q}}\frac{1}{\Big(
\tau +{\bf q}^{2}\Big)\Big(\tau +({\bf q+k})^{2}\Big)}
\nonumber \\
&=& -\frac{g^{2}\tau^{-\varepsilon /2}}{A_{\varepsilon }\varepsilon }\bigg[ \frac{\tau }{1-\varepsilon /2}+\frac{{\bf k}^{2}}{6}+O({\bf k}^{4})\bigg] \, ,
\end{eqnarray}
where $\varepsilon =6-d$ and $A_{\varepsilon }=(4\pi )^{d/2}/\Gamma
(1+\varepsilon /2)$. Renormalization, Eq.(\ref{RenVT}), with $\Gamma
_{N}\rightarrow \mathaccent"7017{\Gamma }_{N}=Z^{-N/2}\Gamma _{N}$ and the identification ${\bf k}^{2}=\hat{K}^{2}$ leads to the $1$-loop result
\begin{eqnarray}
\Gamma _{2}(\hat{K},\tau ) &=&\bigg[ Z_{\tau }-\frac{u}{\varepsilon
(1-\varepsilon /2)}\Big(\mu ^{2}/\tau \Big)^{\varepsilon /2}\bigg] \tau\nonumber \\
 &+&
\bigg[ Z-\frac{u}{6\varepsilon }\Big(\mu ^{2}/\tau \Big)^{\varepsilon /2}
\bigg] \hat{K}^{2}+O(u^{2},\hat{K}^{4})\, .
\end{eqnarray}
The $\varepsilon$-poles are eliminated by choosing minimally
\begin{equation}
Z=1+\frac{u}{6\, \varepsilon }+O(u^{2})\, ,\quad Z_{\tau }=1+\frac{u}{
\varepsilon }+O(u^{2})\, .  \label{RenFac1}
\end{equation}

Now we consider the first order correction $\Gamma _{3}^{1-\text{loop}}$ to the vertex function $\Gamma_3$.
\begin{figure}
\epsfig{file=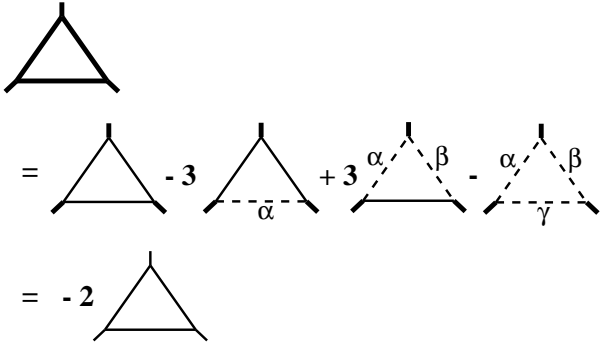,width=8.0cm}
\caption{Reduction of the one-loop vertex diagram.}
\end{figure}
The three-leg diagram constructed of the HRS (bold) propagators, Fig.~2, decomposes in amorphous and liquid (dashed) propagators shown in the second line of Fig.~2. Once more the summation over the colors $\alpha$, $\beta$, and $\gamma$ is implied. We can set $\alpha =\beta =\gamma$ in the diagrams to extract the primitive divergencies. However, then the third and fourth diagram can be discarded because the wave vectors of the propagators of the same color cannot add to a HRS external wave vector at all the vertices. We are left the third line of Fig.~2 where the propagators are replaced by $(\tau +{\bf q}_{i}^{2})^{-1}$. Thus, we find
\begin{equation}
\Gamma_{3}(\{{\bf 0\}})=2g^{3}\int_{{\bf q}}\frac{1}{\big(\tau + {\bf q}^{2}\big)^{3}}=\frac{2g^{3}\tau ^{-\varepsilon /2}}{A_{\varepsilon}\varepsilon }\, .
\end{equation}
After renormalization we get
\begin{equation}
\Gamma _{3}^{1-\text{loop}}(\{0\},\tau )=-g\bigg[ Z_{u}^{1/2}-\frac{2u}{\varepsilon }\Big(\mu ^{2}/\tau \Big)^{\varepsilon /2}+O(u^{2})\bigg] \, .
\end{equation}
It follows the remaining renormalization factor
\begin{equation}
Z_{u}=1+\frac{4u}{\varepsilon }+O(u^{2})\, .  \label{RenFac2}
\end{equation}
All the renormalization factors, Eqs.(\ref{RenFac1},\ref{RenFac2}) are known since a long time as the factors that renormalize the Potts model in the one state limit~\cite{alcantara_80}. Denoting the logarithmic derivatives of the renormalization factors by $\gamma_{...}=\left. \mu \partial \ln Z_{...}/\partial \mu \right| _{0}$, one derives easily the
Gell--Mann-Low-Wilson functions of the RGE as
\begin{eqnarray}
\beta (u) &=&\Big(-\varepsilon +3\gamma -\gamma _{u}\Big)u
\nonumber \\
&=&\Big(-\varepsilon
+\frac{7}{2}u+O(u^{2})\Big)u\, ,
\\  
\gamma (u) &=&-\frac{u}{6}+O(u^{2})\, ,
\\
\kappa_{\tau }(u)&=&\gamma -\gamma_{\tau }=\frac{5u}{6}+O(u^{2})\, .
\end{eqnarray}
The fixed point $u_{\ast }$ as the asymptotic solution of the flow equation $ldu(l)/dl=\beta (u)$ is found to be $u_{\ast }=2\varepsilon /7+O(\varepsilon^{2})$. The critical exponents follow as $\eta =\gamma (u_{\ast})=-\varepsilon /21+O(\varepsilon ^{2})$ and $\nu ^{-1}=2-\kappa _{\tau}(u_{\ast })=2-5\varepsilon /21+O(\varepsilon ^{2})$. These are the known percolation exponents.

\section{The order parameter of the RRN}
\label{ordpar}
In this appendix we discuss some properties of the order parameter $M(\vec{\theta},\tau ,w)$ of the random resistor network above the percolation point. We begin with a mean field consideration based on the saddle point equation steaming from Eq.~(\ref{hamiltonian}):
\begin{eqnarray}
0&=&\frac{\delta {\cal H}_{\text{RRN}}}{\delta \phi({\bf x},\vec{\theta})}=\Big(\tau -\nabla ^{2}-w\nabla _{\theta }^{2}\Big) \phi ({\bf x},\vec{\theta})
\nonumber \\
&-&\frac{g}{2}
\biggl( \phi ({\bf x},\vec{\theta})^{2}-\frac{1}{n}\int_{\vec{\theta}^{\prime }}\phi ( {\bf x},\vec{\theta}^{\prime })^{2}\biggr)\, .  \label{SPEq}
\end{eqnarray}
The last term in this equation follows from the condition that the variation must be done subject to the constraint~(\ref{ConstrRRN}). We seek a spatially homogeneous solution, rotational invariant about $\vec{\theta}=0$ in replica space. Thus, we make the ansatz $\phi ({\bf x},\vec{\theta})=M(\vec{\theta})=m(f(\vec{\theta}^{2})-1)$. Here denotes $f$ a localized function with $\int_{\vec{\theta}^{\prime }}f(\vec{\theta}^{\prime 2})=\int_{\vec{\theta}^{\prime }}1=n$. From Eq.~(\ref{SPEq}) we get, assuming $m\neq 0$ for $\tau <0$,
\begin{eqnarray}
0 &=& \Big(\tau -w\nabla _{\theta }^{2}\Big)f(\vec{\theta}^{2})-\tau 
\nonumber \\
&-&\frac{gm}{2}
\biggl(f(\vec{\theta}^{2})^{2}-2f(\vec{\theta}^{2})+2-\frac{1}{n}\int_{\vec{\theta}^{\prime }}f(\vec{\theta}^{\prime 2})^{2}\biggr)\ .
\end{eqnarray}
We separate this equation in its localized and its delocalized parts:
\begin{eqnarray}
\label{MFEq1}
\Big(\tau -w\nabla _{\theta }^{2}\Big)f(\vec{\theta}^{2})-\frac{gm}{2}\biggl(
f(\vec{\theta}^{2})^{2}-2f(\vec{\theta}^{2})\biggr) &=&0\, ,   
\\
\label{MFEq2}
\tau +\frac{gm}{2}\biggl(2-\frac{1}{n}\int_{\vec{\theta}^{\prime }}f(\vec{\theta}^{\prime 2})^{2}\biggr)  &=&0 \, .  
\end{eqnarray}
Now we write the function $f$ as a Laplace integral
\begin{equation}
f(\vec{\theta}^{2})=\int_{0}^{\infty }dt\,\bar{p}(t)\exp (-t\vec{\theta}^{2})\, .
\end{equation}
Then we have for the term containing the replica space derivative
\begin{eqnarray}
\nabla _{\theta }^{2}f(\vec{\theta}^{2})&=&\int_{0}^{\infty }dt\,\bar{p}(t) \Big(4t^{2}\vec{\theta}^{2}-2tD\Big)\exp (-t\vec{\theta}^{2})
\nonumber \\
&=& \int_{0}^{\infty }dt\,4 \exp (-t\vec{\theta}^{2}) \frac{d}{dt}\Big(t^{2}\bar{p}(t)\Big) \, ,
\end{eqnarray}
where the last equality holds in the limit $D\rightarrow 0$, i.e., $n\rightarrow 1$. Furthermore we can deduce in this limit from the normalization of $f$ that $\int_{0}^{\infty }dt\,\bar{p}(t)=1$ which leads in the limit $n\to 1$ to $\int_{\vec{\theta}}f(\vec{\theta}^{2})^{2}=(\int_{\vec{\theta}}f(\vec{\theta}^{2}))^{2}=1$. Thus, we obtain from Eq.~(\ref{MFEq2}) that the mean field percolation order parameter is given for $\tau <0$ by $m=-2\tau /g=2\left| \tau \right| /g$. Using this result and equating the coefficients of $\exp (-t\vec{\theta}^{2})$ we get finally from Eq.~(\ref{MFEq1})
\begin{equation}
\frac{4w}{\left| \tau \right| }\frac{d}{dt}\Big(t^{2}\bar{p}(t)\Big)=\bar{p}%
(t)-\int_{0}^{t}dt^{\prime }\,\bar{p}(t^{\prime })\bar{p}(t-t^{\prime })\, ,
\end{equation}
which constitutes, after a rescaling, the integro-differential equation~(\ref{DiffGl}). A very good approximative solution of this equation is given by $p(x)=ax^{-2}\exp (-1/4x)$ for $x\ll 1$ and $p(x)=24(bx-3/5)\exp (-bx)$ for $x\gg 1$ with $a=0.56925$ and $b=13.424$~\cite{goldbart_castillo_zippelius_96&zippelius_goldbart_98}. These asymptotic forms yield also very good approximations in the overlapping region $x\approx 1$.

The previous mean field consideration is valid in an exact sense only for dimensions $d\geq 6$. To get information about the behavior of the distribution $p(x)$ on the spatial dimension $d$ below $6$, we calculate the mean macroscopic conductance to infinity $\Sigma_\infty$. We consider spatial length scales large in comparison to the correlation length $\xi$. In this regime the RRN above the percolation point, $\tau <0$, can be considered as a homogeneous material of conductivity $\sigma (\tau )$. Instead of addressing $\Sigma_\infty$ directly we determine the resistance $R_{L}$ of the RRN between in inner sphere of radius $a\gg \xi $ and an outer one of radius $L\gg a$. The solution to this simple problem may be gleaned from many textbooks on electricity and magnetism. One finds that the voltage $V$ behaves as a function of the current $I$ as
\begin{eqnarray}
V &=&\frac{I}{\sigma (d-2)S_{d}}\biggl(\frac{1}{a^{d-2}}-\frac{1}{L^{d-2}} \biggr)=R_{L}I \, ,
\end{eqnarray}
where $S_{d}$ is the surface of the unit sphere in $d$ dimensions. $\Sigma_\infty$ can now be obtained by taking the limit $L\to \infty$. The leading terms in this limit are
\begin{eqnarray}
\Sigma _{L} &=&R_{L}^{-1}\approx \sigma S_{d}\left\{
\begin{array}{cc}
(d-2)a^{d-2} & \text{for }d>2\ , \\
\Big(\ln (L/a)\Big)^{-1} & \text{for }d=2\ , \\
(2-d)L^{-(2-d)} & \text{for }d<2\, .
\end{array}%
\right.   \label{MakKond}
\end{eqnarray}
Hence, the macroscopic conductance to infinity $\Sigma _{\infty }$ is finite in the case of $d>2$ and vanishes for $d\leq 2$. We conclude that the distribution of the conductance to infinity $p(x)$ must develop a $\delta $-peak at $x=0$ if $d\leq 2$ and the order parameter of the RRN vanishes.
If we make a scaling ansatz
\begin{equation}
\Sigma _{L}(\tau )=\left| \tau \right| ^{\phi }F(L/\xi ,a/\xi )
\end{equation}%
and compare with Eq.(\ref{MakKond}), we find the well known exponent $t$ for the macroscopic conductivity $\sigma (\tau )=\left| \tau \right|^{t}$
\begin{equation}
t=(d-2)\nu +\phi
\end{equation}
in all cases.

\end{document}